\documentclass[aps,prl,preprint,superscriptaddress,showpacs]{revtex4}

\usepackage{textcomp}
\usepackage{amsmath}
\usepackage{graphicx}

\newcommand{\layerstruct}[1]{...\textit{#1}...}

\begin{document}

% ******************************
% Title of paper
% ******************************
\title{Zero-Temperature Structures of Atomic Metallic Hydrogen}

% ******************************
% Authors / affiliations
% ******************************
\author{Jeffrey M. McMahon}
\email[]{mcmahonj@illinois.edu}
% \affiliation{The Institute for Condensed Matter Theory, University of Illinois, Urbana-Champaign, IL 61801}
\affiliation{Department of Physics, University of Illinois, Urbana-Champaign, IL 61801}

\author{David M. Ceperley}
\email[]{ceperley@ncsa.uiuc.edu}
\affiliation{Department of Physics, University of Illinois, Urbana-Champaign, IL 61801}
\affiliation{NCSA, University of Illinois, Urbana-Champaign, IL 61801}

% ******************************
% Date
% ******************************
\date{\today}

% ******************************
% Abstract
% ******************************
\begin{abstract}
Ab initio random structure searching with density functional theory was used to determine the zero-temperature structures of atomic metallic hydrogen from $500$ GPa to $5$ TPa. Including zero point motion in the harmonic approximation, we estimate that molecular hydrogen dissociates into a monatomic body-centered tetragonal structure near $500$ GPa ($r_s = 1.225$), which then remains stable to $2.5$ TPa ($r_s = 0.969$). At higher pressures, hydrogen stabilizes in an \layerstruct{ABCABC} planar structure that is remarkably similar to the ground state of lithium, which compresses to the face-centered cubic lattice beyond $5$ TPa ($r_s < 0.86$). At this level of theory, our results provide a complete ab initio description of the atomic metallic structures of hydrogen, resolving one of the most fundamental and long outstanding issues concerning the structures of the elements.
\end{abstract}

% i. I originally included 71.30.+h, but the transition from Cmca to I4_1/amd is a metal-metal transition
% 62.50.-p : High-pressure effects; in solids and liquids
% 64.70.K-, 81.30.-t : Phase transitions; solid-solid
% 64.70.kd : Solid-solid transitions; metals
% 81.30.-t : Solid-solid transitions; phase diagrams of
% 67.80.F- : Solid hydrogen || Quantum solids; solid hydrogen
% 67.80.-s : Quantum crystals
% 71.30.+h : Insulator-metal transitions || Metal-insulator transition

%\keywords{}

% ******************************
% \maketitle must follow title, authors, abstract, \pacs, and \keywords
% ******************************
\maketitle


Since the first prediction of an atomic metallic phase of hydrogen by Wigner and Huntington over 75 years ago \cite{metallic-H_Wigner-JCP-1935}, there have been many theoretical efforts aimed at determining the crystal structures of the zero-temperature phases as a function of density \cite{atomic-H_cryst-struct_Ceperley-PRL-1993, isotrop_H_structs_Ceperley-PRB-1987, isotrop_H_structs_Ashcroft-PRL-1977, ph_H-phases_Cohen-PRL-1989, 9R_sh_H-phases_Cohen-PRB-1991, H_tet-structs_Matsubara-PRB-1997, anisotrop_H-phases_Miyagi-PhysLettA-1989, anisotrop_H-phases_Nagara-JPhysSocJpn-1989}. Such interest is understandable, considering the importance to astrophysics, predictions of high-$T_c$ superconductivity \cite{high-Tc_SC_H_Ashcroft-PRL-1968}, and the possibility of a low or even zero-temperature metallic liquid \cite{quant-liq_0K_H_Bonex-Nat-1968}. Despite the importance and corresponding efforts, there is still no conclusive understanding of this most basic and fundamental aspect. These efforts have been hindered by the fact that experiments have only been able to reach pressures of just over $300$ GPa \cite{solid_H_exp_highest-P_Ruoff-Nature-1998}, which is lower than that of the atomic phase(s).

Previous studies have taken the approach of simply proposing candidate structures, leading to diverse predictions. In some cases, isotropic structures have been predicted as the ground state \cite{atomic-H_cryst-struct_Ceperley-PRL-1993, isotrop_H_structs_Ceperley-PRB-1987, isotrop_H_structs_Ashcroft-PRL-1977, H_tet-structs_Matsubara-PRB-1997}, while in others anisotropic ones have \cite{ph_H-phases_Cohen-PRL-1989, 9R_sh_H-phases_Cohen-PRB-1991, anisotrop_H-phases_Miyagi-PhysLettA-1989, anisotrop_H-phases_Nagara-JPhysSocJpn-1989, anisotrop_planar_structs_0P_Kholas-SovPhysJETP-1972, anisotrop_planar_structs_ZPE_Peffects_Kholas-SovPhysJETP-1972}. This diversity is related to the primary disadvantage of such an approach, in that only a few select structures can be considered at any one time. It is apparent that, in analogy with the other alkali metals, even an elemental solid can have a rather complex structure \cite{low-T_cryst-struct_Li_Overhauser-PRL-1984}. Recently, however, more robust methods for determining crystal structures have been proposed, such as the ab initio random structure searching (AIRSS) method by Pickard and Needs \cite{AIRSS_orig_Needs-PRL-2006}. In this approach, a number of random configurations are each relaxed to the ground state at constant pressure. After enough trials, a good sampling of the atomic configuration space is obtained and the ground state structure is generated with a high probability. Such an approach has been used to accurately predict a number of structures, including more complicated ones than considered here, such as silane \cite{AIRSS_orig_Needs-PRL-2006} and the highest pressure molecular phase of hydrogen, phase III \cite{AIRSS_H2-phaseIII_Needs-NatPhys-2007}.

% [JMM: it would be nice to estimate the DMC vs DFT error]
In this Letter, we use AIRSS to determine the zero-temperature structures of atomic metallic hydrogen. Our calculations were performed using the Quantum Espresso ab initio density functional theory (DFT) code \cite{QE-2009}. A norm-conserving Troullier--Martins pseudopotential \cite{TM-PP_Troullier-Martins-PRB-1991} with a cutoff radius of $0.5$ a.u.\ was used to replace the true $1/r$ Coulomb potential of hydrogen, along with the Perdew, Burke, and Ernzerhof exchange and correlation functional \cite{PBE_exch-correl_PRL-1996}. 
%(The errors incurred by both approximations were found to be small by performing a few static-lattice diffusion Monte Carlo calculations.) 
% [JMM: it would be nice to estimate the DMC vs DFT error, but we can only make a rough comparison, and I don't think the DMC results are free of finite-size effects] For example, at $r_s = 1$ our DFT calculations predict a difference in energy of -4.269 mRy/proton between $I4_1/amd$ ($c/a > 1$) and $I4_1/amd$ ($c/a < 1$), whereas DMC predicts -2.436 mRy/proton.
A basis set of plane-waves with a cutoff of $2721$ eV was used for the random structure searching, and then increased to $2993$ eV for recalculating detailed enthalpy vs.\ pressure curves. For Brillouin-zone sampling, $14 \times 14 \times 14$ \textit{k}-points were used for the random searching and then significantly increased for recalculating enthalpy curves. While both the plane-wave cutoff and number of \textit{k}-points may seem exceedingly high, such values were found necessary to ensure convergence of each structure to better than $1.5$ meV/proton in energy and the pressure to better than $0.1$ GPa/proton. 
%(Due to error cancellations, relative comparisons between structures should therefore be possible to even less.)
Phonons were calculated using density functional perturbation theory. Typical relaxations included at least $100$ random structures at each pressure considered. In most cases this appeared to be enough to generate the low-enthalpy structure(s) multiple times. At pressures where the results were considered inconclusive (e.g., a low-enthalpy structure generated only once or twice), additional relaxations were performed. Note that herein we will refer to each structure by its Hermann--Mauguin space-group symbol (international notation), but will also provide more common lattice names, where applicable.

\begin{figure}
  \includegraphics[scale=0.28, bb=0 0 877 847]{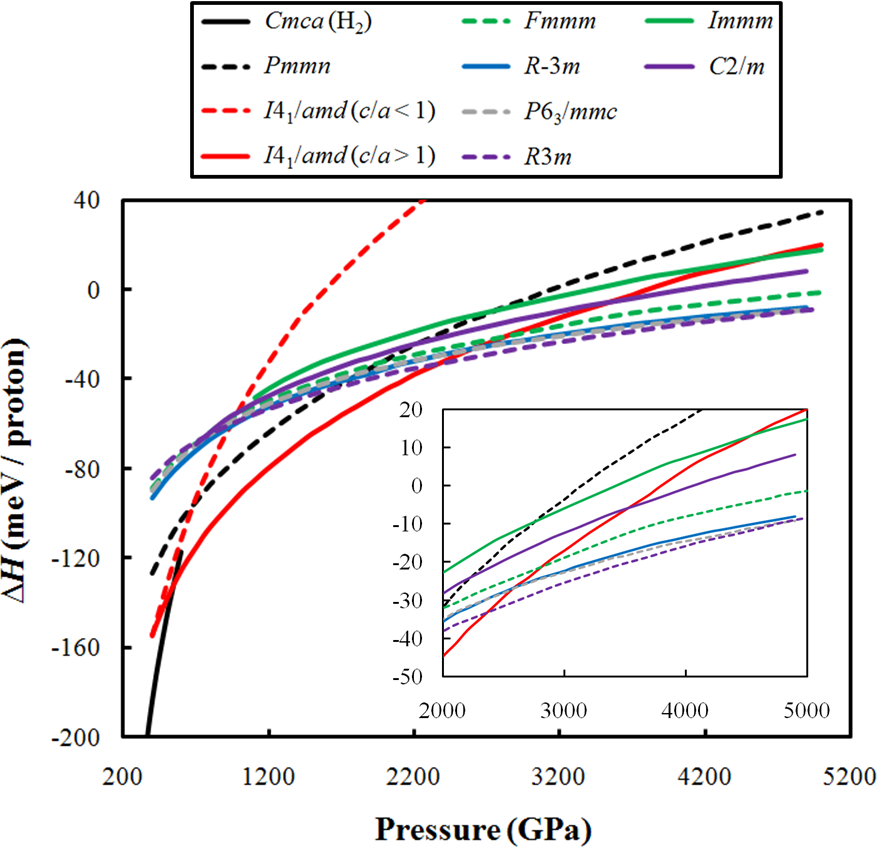}
  % H_no-ZPE_top-legend.png: 877x847 pixel, 72dpi, 30.94x29.88 cm, bb=0 0 877 847
  \caption{(color online). Zero-temperature enthalpies of the crystal structures of atomic metallic hydrogen, not including proton ZPE. The inset shows an expanded view of the ultrahigh pressure region.}
  \label{fig:phase-diag}
\end{figure}

% [JMM: I cannot remember if I did 6 atom unit cell relaxations at 4.5 TPa]
AIRSS was first carried out for unit cells containing $4$ and $6$ atoms at pressures from $500$ GPa to $4.5$ TPa in intervals of $500$ GPa. Such relaxations implicitely include searches over unit cells of their factors -- i.e., those with $1$, $2$, or $3$ atoms. While structures with unit cells of $5$ or more than $7$ atoms are certainly possible, they are unlikely to occur based on comparisons with other elemental structures. For example, lithium, the closest element to monatomic hydrogen, has a ground state structure consisting of a $3$ atom rhombohedral unit cell with space-group $R$-$3m$ \cite{low-T_cryst-struct_Li_Overhauser-PRL-1984}. Detailed enthalpy vs.\ pressure curves were then calculated for each structure found, giving the results in Fig.\ \ref{fig:phase-diag}.
Note that the enthalpy $H$ shown is relative to the face-centered cubic (fcc) lattice (space-group $Fm$-3$m$). Also note that the body-centered cubic (bcc) lattice (space-group $Im$-$3m$), which was assumed to be the structure of dense hydrogen originally proposed by Wigner and Huntington \cite{metallic-H_Wigner-JCP-1935}, is not shown in Fig.\ \ref{fig:phase-diag} for clarity, but it is less stable than fcc by approximately $11$ meV/proton over the entire pressure range considered [which increases by a further $60$ -- $70$ meV/proton above $1$ TPa when proton zero-point energy (ZPE) is included, as discussed below].
% JMM: note that the 1 TPa is an estimation, 60 -- 70 meV/proton is taken from 1.5 TPa+ data, as I don't have 1 TPa data. 
% Although they were not generated during our searches, we also include the structures of diamond (space-group $Fd$-3$m$) and $\beta$-$Sn$ (space-group $I4_1/amd$ with a $c / a$ ratio less than unity), which are often considered good candidates for the atomic metallic phase of hydrogen \cite{H_tet-structs_Matsubara-PRB-1997, atomic-H_cryst-struct_Ceperley-PRL-1993, 9R_sh_H-phases_Cohen-PRB-1991}.

\begin{figure}
  \includegraphics[scale=0.19, bb=0 0 1291 764]{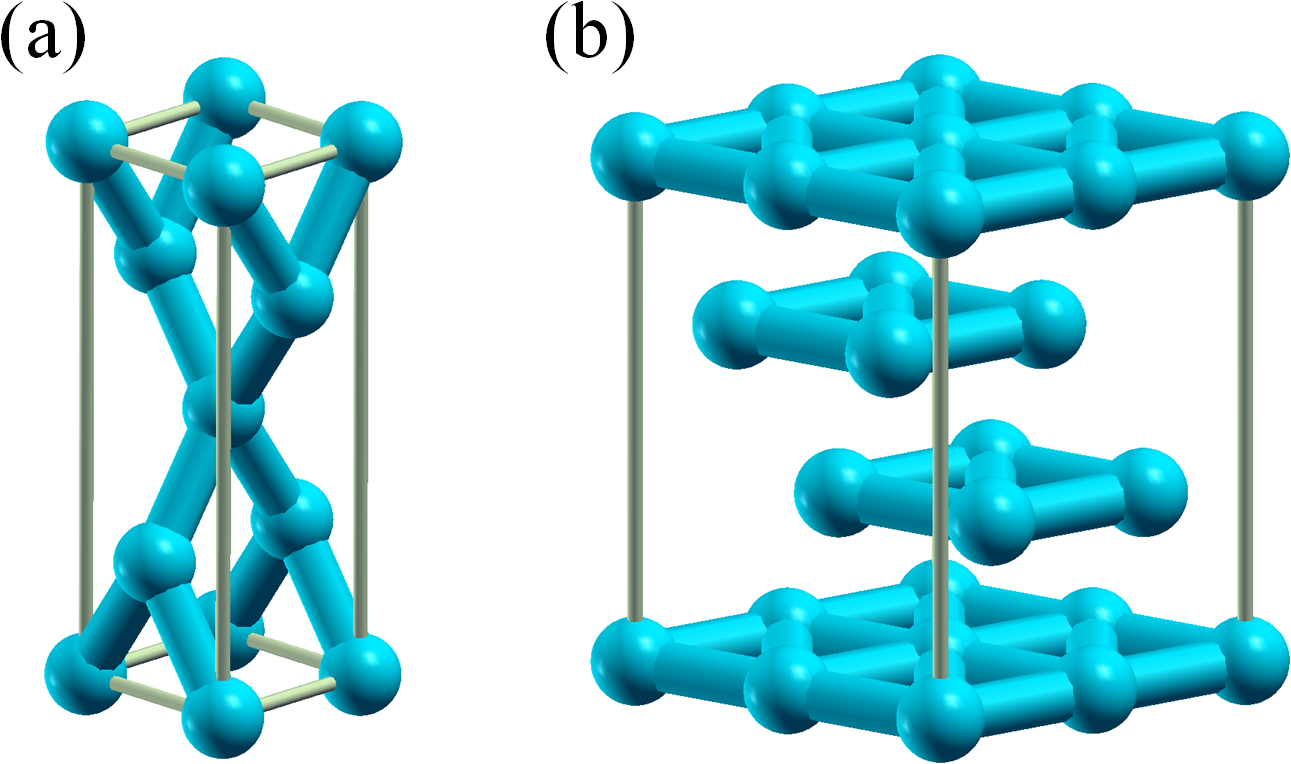}
  % I41amd-1p5TPa_R-3m-2p5TPa.png: 1291x764 pixel, 72dpi, 45.54x26.95 cm, bb=0 0 1291 764
  \caption{(color online). Structures of the most stable zero-temperature phases of atomic metallic hydrogen. (a) Unit cell of $I4_1/amd$ ($c/a > 1$) at $1.5$ TPa. (b) $2 \times 2 \times 1$ supercell of $R$-$3m$ at $3.5$ TPa. Ficticious bonds have been drawn for clarity.}
  \label{fig:struct_diagrams}
\end{figure}

% i. the r_s value here was taken from I4_1amd at 500 GPa
% [JMM: note that I removed the references to Cs-IV and b-Sn, since I don't discuss what structures the other space groups go to]
Given that our calculations span a large range in pressures, and also much higher than previously considered, Fig.\ \ref{fig:phase-diag} contains a significant number of structures. Below $500$ GPa ($r_s = 1.225$), the most stable structure is the molecular phase $Cmca$, which has previously been predicted by both theoretical calculations \cite{metallic_H2_structs_Ashcroft-Nat-2000} and AIRSS \cite{AIRSS_H2-phaseIII_Needs-NatPhys-2007}. Near $500$ GPa, $Cmca$ dissociates into a monatomic body-centered tetragonal structure of space-group $I4_1/amd$ with a $c/a$ ratio greater than unity (e.g., $c/a = 2.588$ at $500$ GPa), which is shown in Fig.\ \ref{fig:struct_diagrams}(a).
%
% [JMM: I am not sure why I decided to use the structure at 3.5 TPa, as opposed to 3 TPa which is often used for other high pressure references.]
This transition is also consistent with previous calculations \cite{AIRSS_H2-phaseIII_Needs-NatPhys-2007}. Our searches also generated a corresponding $I4_1/amd$ structure with a $c/a$ ratio less than unity (e.g., $c/a = 0.877$ at $500$ GPa). However, while both structures are similar in enthalpy near $500$ GPa, the latter quickly becomes much less stable with an increase in pressure.
%, and the precise physics associated with it represent an intriguing problem in itself \cite{quant-liq_0K_H_Bonex-Nat-1968}.

\begin{figure}
  \includegraphics[scale=0.15, bb=0 0 1633 850]{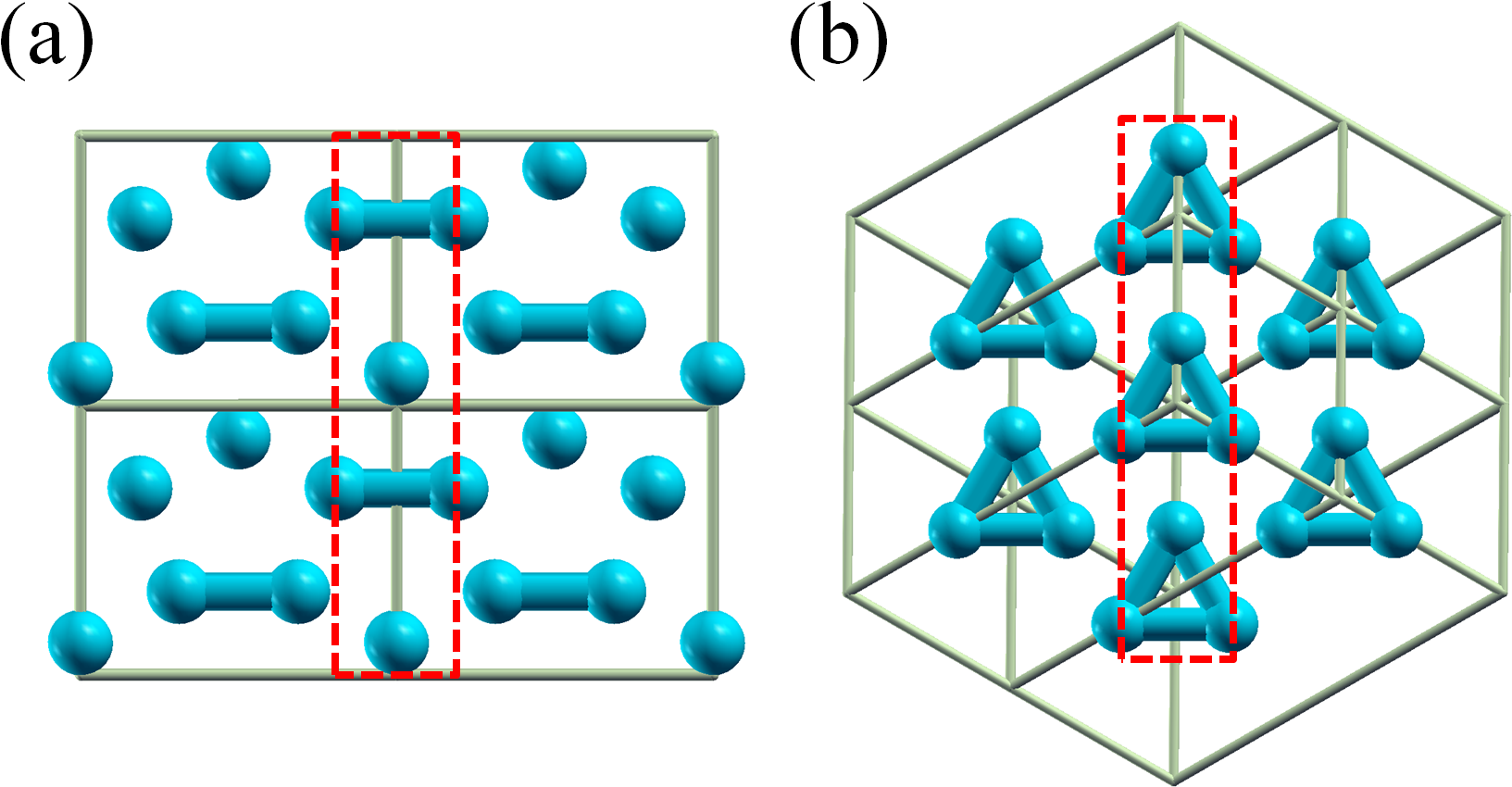}
  % Pmmn-2TPa_R3m-3TPa.png: 1633x850 pixel, 72dpi, 57.61x29.99 cm, bb=0 0 1633 850
  \caption{(color online). Structures of (a) $Pmmn$ at $1.5$ TPa and (b) $R3m$ at $3$ TPa. Note that the unit cells shown in (b) are actually $2 \times 1 \times 1$ supercells. The dotted lines border linear chains, which stick out of the plane relative to their neighbors.}
  \label{fig:Pmmn_R3m}
\end{figure}

% i. the r_s value here was taken from I4_1amd at 2.5 TPa
$I4_1/amd$ is found to remain stable until approximately $2.5$ TPa ($r_s = 0.969$), resisting compression along the $c$ axis (e.g., $c/a = 2.993$ at $2.5$ TPa). This result is similar to the conclusion of a previous study that considered a family of tetragonal structures \cite{H_tet-structs_Matsubara-PRB-1997}. The only other structure close in enthalpy to $I4_1/amd$ generated during our searches was $Pmmn$, which is still $15$ meV/proton less stable over the entire pressure range considered. While relatively unstable, $Pmmn$ does form an intriguing structure. Below $1.9$ TPa it is monatomic. However, with increasing pressure a pairing between some atoms occurs forming a mix of molecular and atomic hydrogen that arranges in linear chains; see Fig.\ \ref{fig:Pmmn_R3m}(a).

Near $2.5$ TPa ($r_s = 0.969$) four additional structures with similar enthalpies become important. The least stable is a face-centered orthorhombic structure with space-group $Fmmm$, which is similar to the fcc latice except elongated along both the $b$ and $c$ axes (e.g., $b/a = 1.724$ and $c/a = 2.021$ at $3$ TPa). Both structures are planar, however $Fmmm$ stacks in the sequence \layerstruct{ABAB} whereas fcc does so as \layerstruct{ABCABC} (and is also close-packed). Slightly more stable by $5$ -- $6$ meV/proton are two structures nearly equal in enthalpy. One is $P6_3/mmc$, a hexagonal structure that is also planar with \layerstruct{ABAB} stacking, and is similar to the hexagonal close-packed (hcp) lattice except elongated along the $c$ axis (e.g., $c/a = 2.008$ at $3$ TPa). Note that hcp is also of space-group $P6_3/mmc$ with $c/a = 1.633$, and is not shown in Fig.\ \ref{fig:phase-diag} for clarity but is more stable than fcc by about $4.5$ meV/proton over the entire pressure range considered (although becomes less stable by $30$ -- $47$ meV/proton when ZPE is included, as discussed below). The other structure is $R$-3$m$, a planar structure with \layerstruct{ABCABC} stacking (and with $c/a = 3.028$ at $3$ TPa, for example), as shown in Fig.\ \ref{fig:struct_diagrams}(b). (Recall that fcc forms a close-packed version of this stacking sequence, as discussed above.) The most stable structure of this group (by a further $2$ -- $3$ meV/proton at $3.5$ TPa, for example) is $R3m$, which is formed from a rhombohedral unit cell consisting of triatomic molecules; see Fig.\ \ref{fig:Pmmn_R3m}(b). This structure is likely derived from $Pmmn$ by a compression of the linear chains, followed by a slight distortion.

It is interesting to note that three out of the four structures in this pressure range are anisotropic and planar. This is consistent with previous work that predicted such structures with an increase in density \cite{anisotrop_H-phases_Miyagi-PhysLettA-1989, anisotrop_H-phases_Nagara-JPhysSocJpn-1989}. Similar structures have even been proposed at zero pressure \cite{anisotrop_planar_structs_0P_Kholas-SovPhysJETP-1972}. Perhaps more interesting is that $R$-$3m$ is also the space-group of the ground state of lithium \cite{low-T_cryst-struct_Li_Overhauser-PRL-1984}, the structure of which, termed $9R$, differs primarily by its stacking sequence (lithium stacks as \layerstruct{ABCBCACAB}) and packing efficiency. In fact, the $9R$ structure has been previously suggested as a good candidate for the atomic metallic phase of hydrogen \cite{9R_sh_H-phases_Cohen-PRB-1991}. Relative to $R$-$3m$, however, $9R$ is unstable. This can be inferred by comparing their relative stabilities to bcc. $9R$ is predicted to transform to bcc at $1090 \pm 100$ GPa \cite{9R_sh_H-phases_Cohen-PRB-1991}, while $R$-$3m$ is stable even above $5$ TPa (as it is more stable than fcc; see Fig.\ \ref{fig:phase-diag}). 

Two additional structures with significantly higher enthalpy were also generated during our searches in the pressure range $2$ -- $3$ TPa, $Immm$ and $C2/m$ (which are qualitatively very similar). Much like $R3m$, these structures are likely derived from $Pmmn$. Below $3$ TPa they are very similar to $Pmmn$, comprised of chains of atomic and molecular hydrogen. However, their behavior with increasing pressure is different. $Pmmn$ compresses both along and between the linear chains, forming triatomic molecules connected to their counterparts in neighboring planes. $Immm$ and $C2/m$, on the other hand, resist compression along the $c$ axis, forming molecular chains. It should be noted that such structures have recently been suggested as the ultrahigh pressure ground state of metallic hydrogen \cite{triatomic_H_structs_Chinese-arxiv-2010}.

% i. note that the r_s value for above 5 TPa was taken from fcc at ~52 Mbar
By approximately $4$ TPa our searches began generating the ``simple'' lattices, such as fcc. This suggests that the entire pressure range from $500$ GPa to $5$ TPa is well mapped out, with the result that the molecular phase ($Cmca$) dissociates into a body-centered tetragonal structure ($I4_1/amd$) near $500$ GPa ($r_s = 1.225$), transforming to a planar ($R$-$3m$ or $P6_3/mmc$) or triatomic ($R3m$) structure near $2.5$ TPa ($r_s = 0.969$), which either compresses or transforms to a close-packed lattice (fcc or hcp) above $5$ TPa (approximately $r_s < 0.86$). 

% [JMM: the description of the 0K phase-diagram by adding ZPE to the diagram in Fig. 1 is a little confusing, but I'll leave that description for now.]
The scenario outlined above is based on lattices of infinitely massive protons. It therefore actually describes the isotopes of hydrogen with heavier nuclei, including tritium and possibly deuterium. The light proton mass, however, causes hydrogen to have a large ZPE, which must be estimated in order to determine the correct ordering of the structures. 
%We considered structures within $\sim \! 20$ meV/proton in Fig.\ \ref{fig:phase-diag} for further calculation. 
Accurately estimating ZPE via DFT is particularly challenging because proton zero point motion is anharmonic in atomic metallic hydrogen \cite{atomic-H_cryst-struct_Ceperley-PRL-1993}. A full analysis of these effects is beyond the scope of this work, but some insight can nonetheless be obtained by treating the motion as harmonic.

Phonons were calculated at the $\Gamma$ point for each structure and the ZPE was estimated in the harmonic approximation: $E_\text{ZPE} = (1/2) \hbar \langle \omega_\text{ph} \rangle$, where $\langle \omega_\text{ph} \rangle$ is the average phonon frequency (which does not include the zero-frequency acoustic modes).
\begin{figure}
  \includegraphics[scale=0.28, bb=0 0 877 845]{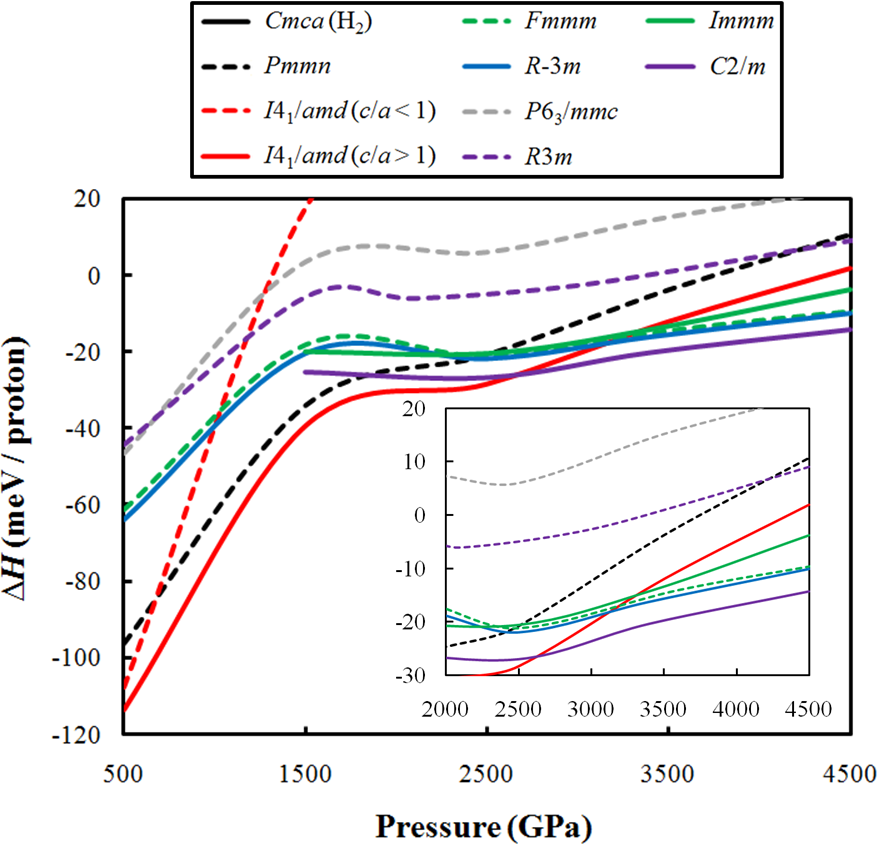}
  % H_ZPE_top-legend_concise-range.png: 877x845 pixel, 72dpi, 30.94x29.81 cm, bb=0 0 877 845
  \caption{(color online). Zero-temperature enthalpies of the crystal structures of atomic metallic hydrogen, including proton ZPE. The inset shows an expanded view of the ultrahigh pressure region. Note that a single point representing the molecular phase $Cmca$ at $500$ GPa is off the scale of the figure, with a relative enthalpy of approximately $-132$ meV/proton.}
  \label{fig:phase-diag_w-phonons}
\end{figure}
Calculating this energy from $500$ GPa to $4.5$ TPa in intervals of $500$ GPa and adding it to the enthalpy in Fig.\ \ref{fig:phase-diag} gives the modified zero-temperature phase diagram shown in Fig.\ \ref{fig:phase-diag_w-phonons}. The relative stabilities of the structures are found to drastically change, as expected. For example, while $I4_1/amd$ remains the ground state from $500$ GPa to $2.5$ TPa, $Pmmn$ becomes much closer in enthalpy. 
% [JMM: Ceperley asks to explain further the following comment; In fact, I am going to take it out because I do not necessarily believe it as R3m which is actually triatomic increases in energy] The lower ZPE of $Pmmn$ relative to $I4_1/amd$ is understandable, considering that the partially covalent bonds of the triatomic molecules can stabilize the protons against ZPM. 
At higher pressures, the near degeneracy of $Fmmm$, $P6_3/mmc$, $R$-3$m$, and $R3m$ is lifted. $R$-3$m$ becomes the clear ground state of this group, followed by $R3m$ and $P6_3/mmc$ at much higher enthalpy, while $Fmmm$ appears especially unstable. This is not to say that $R$-$3m$ is necessarily the ground state, as $Immm$ and $C2/m$ become comparable in enthalpy.
%, which is again understandable based on the same rationale. 
It is interesting to note that both \layerstruct{ABAB} planar structures ($Fmmm$ and $P6_3/mmc$) become relatively unstable, while the \layerstruct{ABCABC} one ($R$-$3m$) does not, considering that the major difference is the stacking sequence. Before leaving this topic, it is again worth stressing that proton zero point motion is anharmonic \cite{atomic-H_cryst-struct_Ceperley-PRL-1993}, and therefore definitive conclusions regarding the ordering of the structures should not be made even on the basis of Fig.\ \ref{fig:phase-diag_w-phonons}.

Because such a large pressure range was considered, it is possible that our searches could have missed some structures. An indication of this as well as possible structure instabilities is given by the appearance of imaginary phonon frequencies.
% [JMM] The energies of these frequencies, again in the harmonic approximation, are shown in Fig.\ \ref{fig:instabilities} 
However, it turns out that most of the structures do not exhibit such frequencies over their respective ranges of importance (e.g., $I4_1/amd$ and $Pmmn$ below approximately $2.5$ TPa). Although, both $Immm$ and $C2/m$, which become comparable in enthalpy to $R$-$3m$ when proton ZPE is included (Fig.\ \ref{fig:phase-diag_w-phonons}), are unstable by $2$ -- $6.5$ meV/proton (also estimated in the harmonic approximation) over the entire pressure range considered, while $R$-$3m$ is completely stable. This suggests that $R$-$3m$ is the ultrahigh pressure ground state structure (within the harmonic approximation, as discussed above).

In summary, we performed AIRSS to determine the zero-temperature structures of atomic metallic hydrogen at pressures from $500$ GPa to $5$ TPa. We estimate that molecular hydrogen dissociates into a monatomic body-centered tetragonal structure near $500$ GPa ($r_s = 1.225$), which then remains stable to $2.5$ TPa ($r_s = 0.969$). At higher pressures, the most stable phase becomes a hexagonal \layerstruct{ABCABC} planar structure which is similar to the low-temperature ground state of lithium. With increasing pressure this structure compresses, likely with a continuous change to fcc above $5$ TPa ($r_s < 0.86$), analogous to the situation proposed in Ref.\ \cite{anisotrop_planar_structs_ZPE_Peffects_Kholas-SovPhysJETP-1972}. An estimation of the proton ZPE was given in the harmonic approximation. However, a more detailed study including anharmonic effects is necessary to determine the precise ordering of the structures. The fact that we found a significant number of structures at zero-temperature with similar enthalpies suggests that most (if not all) of them will become important at finite temperature, when entropic effect are relevant. A study of the effects of both anharmonicity and temperature is currently underway using coupled electron--ion Monte Carlo \cite{CEIMC_Ceperley-LecNotPhys-2006}.
%Nonetheless, our results provide a finite and complete set of structures that comprise the phases of atomic metallic hydrogen. 

% ******************************
% ACKNOWLEDGMENTS
% ******************************

\begin{acknowledgments}
J.M.M. and D.M.C. were supported by DOE DE-FC02-06ER25794 and DE-FG52-09NA29456. We thank Jeremy McMinis for performing quantum Monte Carlo simulations to validate our use of pseudopotential and exchange and correlation functional. We also thank Carlo Pierleoni and Miguel Morales for helpful discussions.
\end{acknowledgments}

%\bibliography{../Bibtex_Refs/journal_names_s,../Bibtex_Refs/cond_matt_phys/hydrogen_cryst-struct,../Bibtex_Refs/cond_matt_phys/hydrogen_quant-fluid,../Bibtex_Refs/cond_matt_phys/alkali_metals,../Bibtex_Refs/comp_phys/comp_phys}

\begin{thebibliography}{22}
\expandafter\ifx\csname natexlab\endcsname\relax\def\natexlab#1{#1}\fi
\expandafter\ifx\csname bibnamefont\endcsname\relax
  \def\bibnamefont#1{#1}\fi
\expandafter\ifx\csname bibfnamefont\endcsname\relax
  \def\bibfnamefont#1{#1}\fi
\expandafter\ifx\csname citenamefont\endcsname\relax
  \def\citenamefont#1{#1}\fi
\expandafter\ifx\csname url\endcsname\relax
  \def\url#1{\texttt{#1}}\fi
\expandafter\ifx\csname urlprefix\endcsname\relax\def\urlprefix{URL }\fi
\providecommand{\bibinfo}[2]{#2}
\providecommand{\eprint}[2][]{\url{#2}}

\bibitem[{\citenamefont{Wigner and
  Huntington}(1935)}]{metallic-H_Wigner-JCP-1935}
\bibinfo{author}{\bibfnamefont{E.}~\bibnamefont{Wigner}} \bibnamefont{and}
  \bibinfo{author}{\bibfnamefont{H.~B.} \bibnamefont{Huntington}},
  \bibinfo{journal}{J. Chem. Phys.} \textbf{\bibinfo{volume}{3}},
  \bibinfo{pages}{1748} (\bibinfo{year}{1935}).

\bibitem[{\citenamefont{Natoli et~al.}(1993)\citenamefont{Natoli, Martin, and
  Ceperley}}]{atomic-H_cryst-struct_Ceperley-PRL-1993}
\bibinfo{author}{\bibfnamefont{V.}~\bibnamefont{Natoli}},
  \bibinfo{author}{\bibfnamefont{R.~M.} \bibnamefont{Martin}},
  \bibnamefont{and} \bibinfo{author}{\bibfnamefont{D.~M.}
  \bibnamefont{Ceperley}}, \bibinfo{journal}{Phys. Rev. Lett.}
  \textbf{\bibinfo{volume}{70}}, \bibinfo{pages}{1952} (\bibinfo{year}{1993}).

\bibitem[{\citenamefont{Barbee~III et~al.}(1989)\citenamefont{Barbee~III,
  Garc\'{\i}a, Cohen, and Martins}}]{ph_H-phases_Cohen-PRL-1989}
\bibinfo{author}{\bibfnamefont{T.~W.} \bibnamefont{Barbee~III}},
  \bibinfo{author}{\bibfnamefont{A.}~\bibnamefont{Garc\'{\i}a}},
  \bibinfo{author}{\bibfnamefont{M.~L.} \bibnamefont{Cohen}}, \bibnamefont{and}
  \bibinfo{author}{\bibfnamefont{J.~L.} \bibnamefont{Martins}},
  \bibinfo{journal}{Phys. Rev. Lett.} \textbf{\bibinfo{volume}{62}},
  \bibinfo{pages}{1150} (\bibinfo{year}{1989}).

\bibitem[{\citenamefont{Ebina and
  Miyagi}(1989)}]{anisotrop_H-phases_Miyagi-PhysLettA-1989}
\bibinfo{author}{\bibfnamefont{K.}~\bibnamefont{Ebina}} \bibnamefont{and}
  \bibinfo{author}{\bibfnamefont{H.}~\bibnamefont{Miyagi}},
  \bibinfo{journal}{Phys. Lett. A} \textbf{\bibinfo{volume}{142}},
  \bibinfo{pages}{237} (\bibinfo{year}{1989}).

\bibitem[{\citenamefont{Ceperley and
  Alder}(1987)}]{isotrop_H_structs_Ceperley-PRB-1987}
\bibinfo{author}{\bibfnamefont{D.~M.} \bibnamefont{Ceperley}} \bibnamefont{and}
  \bibinfo{author}{\bibfnamefont{B.~J.} \bibnamefont{Alder}},
  \bibinfo{journal}{Phys. Rev. B} \textbf{\bibinfo{volume}{36}},
  \bibinfo{pages}{2092} (\bibinfo{year}{1987}).

\bibitem[{\citenamefont{Straus and
  Ashcroft}(1977)}]{isotrop_H_structs_Ashcroft-PRL-1977}
\bibinfo{author}{\bibfnamefont{D.~M.} \bibnamefont{Straus}} \bibnamefont{and}
  \bibinfo{author}{\bibfnamefont{N.~W.} \bibnamefont{Ashcroft}},
  \bibinfo{journal}{Phys. Rev. Lett.} \textbf{\bibinfo{volume}{38}},
  \bibinfo{pages}{415} (\bibinfo{year}{1977}).

\bibitem[{\citenamefont{Barbee~III and
  Cohen}(1991)}]{9R_sh_H-phases_Cohen-PRB-1991}
\bibinfo{author}{\bibfnamefont{T.~W.} \bibnamefont{Barbee~III}}
  \bibnamefont{and} \bibinfo{author}{\bibfnamefont{M.~L.} \bibnamefont{Cohen}},
  \bibinfo{journal}{Phys. Rev. B} \textbf{\bibinfo{volume}{44}},
  \bibinfo{pages}{11563} (\bibinfo{year}{1991}).

\bibitem[{\citenamefont{Nagao et~al.}(1997)\citenamefont{Nagao, Nagara, and
  Matsubara}}]{H_tet-structs_Matsubara-PRB-1997}
\bibinfo{author}{\bibfnamefont{K.}~\bibnamefont{Nagao}},
  \bibinfo{author}{\bibfnamefont{H.}~\bibnamefont{Nagara}}, \bibnamefont{and}
  \bibinfo{author}{\bibfnamefont{S.}~\bibnamefont{Matsubara}},
  \bibinfo{journal}{Phys. Rev. B} \textbf{\bibinfo{volume}{56}},
  \bibinfo{pages}{2295} (\bibinfo{year}{1997}).

\bibitem[{\citenamefont{Nagara}(1989)}]{anisotrop_H-phases_Nagara-JPhysSocJpn-%
1989}
\bibinfo{author}{\bibfnamefont{H.}~\bibnamefont{Nagara}}, \bibinfo{journal}{J.
  Phys. Soc. Jpn.} \textbf{\bibinfo{volume}{58}}, \bibinfo{pages}{3861}
  (\bibinfo{year}{1989}).

\bibitem[{\citenamefont{Ashcroft}(1968)}]{high-Tc_SC_H_Ashcroft-PRL-1968}
\bibinfo{author}{\bibfnamefont{N.~W.} \bibnamefont{Ashcroft}},
  \bibinfo{journal}{Phys. Rev. Lett.} \textbf{\bibinfo{volume}{21}},
  \bibinfo{pages}{1748} (\bibinfo{year}{1968}).

\bibitem[{\citenamefont{Bonev et~al.}(2004)\citenamefont{Bonev, Schwegler,
  Ogitsu, and Galli}}]{quant-liq_0K_H_Bonex-Nat-1968}
\bibinfo{author}{\bibfnamefont{S.~A.} \bibnamefont{Bonev}},
  \bibinfo{author}{\bibfnamefont{E.}~\bibnamefont{Schwegler}},
  \bibinfo{author}{\bibfnamefont{T.}~\bibnamefont{Ogitsu}}, \bibnamefont{and}
  \bibinfo{author}{\bibfnamefont{G.}~\bibnamefont{Galli}},
  \bibinfo{journal}{Nature} \textbf{\bibinfo{volume}{431}},
  \bibinfo{pages}{669} (\bibinfo{year}{2004}).

\bibitem[{\citenamefont{Narayana et~al.}(1998)\citenamefont{Narayana, Luo,
  Orloff, and Ruoff}}]{solid_H_exp_highest-P_Ruoff-Nature-1998}
\bibinfo{author}{\bibfnamefont{C.}~\bibnamefont{Narayana}},
  \bibinfo{author}{\bibfnamefont{H.}~\bibnamefont{Luo}},
  \bibinfo{author}{\bibfnamefont{J.}~\bibnamefont{Orloff}}, \bibnamefont{and}
  \bibinfo{author}{\bibfnamefont{A.~L.} \bibnamefont{Ruoff}},
  \bibinfo{journal}{Nature} \textbf{\bibinfo{volume}{393}}, \bibinfo{pages}{46}
  (\bibinfo{year}{1998}).

\bibitem[{\citenamefont{Brovman et~al.}(1972)\citenamefont{Brovman, Kagan, and
  Kholas}}]{anisotrop_planar_structs_0P_Kholas-SovPhysJETP-1972}
\bibinfo{author}{\bibfnamefont{E.~G.} \bibnamefont{Brovman}},
  \bibinfo{author}{\bibfnamefont{Y.}~\bibnamefont{Kagan}}, \bibnamefont{and}
  \bibinfo{author}{\bibfnamefont{A.}~\bibnamefont{Kholas}},
  \bibinfo{journal}{Sov. Phys. JETP} \textbf{\bibinfo{volume}{34}},
  \bibinfo{pages}{1300} (\bibinfo{year}{1972}).

\bibitem[{\citenamefont{Kagan et~al.}(1977)\citenamefont{Kagan, Pushkarev, and
  Kholas}}]{anisotrop_planar_structs_ZPE_Peffects_Kholas-SovPhysJETP-1972}
\bibinfo{author}{\bibfnamefont{Y.}~\bibnamefont{Kagan}},
  \bibinfo{author}{\bibfnamefont{V.~V.} \bibnamefont{Pushkarev}},
  \bibnamefont{and} \bibinfo{author}{\bibfnamefont{A.}~\bibnamefont{Kholas}},
  \bibinfo{journal}{Sov. Phys. JETP} \textbf{\bibinfo{volume}{46}},
  \bibinfo{pages}{511} (\bibinfo{year}{1977}).

\bibitem[{\citenamefont{Overhauser}(1984)}]{low-T_cryst-struct_Li_Overhauser-P%
RL-1984}
\bibinfo{author}{\bibfnamefont{A.~W.} \bibnamefont{Overhauser}},
  \bibinfo{journal}{Phys. Rev. Lett.} \textbf{\bibinfo{volume}{53}},
  \bibinfo{pages}{64} (\bibinfo{year}{1984}).

\bibitem[{\citenamefont{Pickard and Needs}(2006)}]{AIRSS_orig_Needs-PRL-2006}
\bibinfo{author}{\bibfnamefont{C.~J.} \bibnamefont{Pickard}} \bibnamefont{and}
  \bibinfo{author}{\bibfnamefont{R.~J.} \bibnamefont{Needs}},
  \bibinfo{journal}{Phys. Rev. Lett.} \textbf{\bibinfo{volume}{97}},
  \bibinfo{pages}{045504} (\bibinfo{year}{2006}).

\bibitem[{\citenamefont{Pickard and
  Needs}(2007)}]{AIRSS_H2-phaseIII_Needs-NatPhys-2007}
\bibinfo{author}{\bibfnamefont{C.~J.} \bibnamefont{Pickard}} \bibnamefont{and}
  \bibinfo{author}{\bibfnamefont{R.~J.} \bibnamefont{Needs}},
  \bibinfo{journal}{Nature Phys.} \textbf{\bibinfo{volume}{3}},
  \bibinfo{pages}{473} (\bibinfo{year}{2007}).

\bibitem[{\citenamefont{Giannozzi et~al.}(2009)\citenamefont{Giannozzi, Baroni,
  Bonini, Calandra, Car, Cavazzoni, Ceresoli, Chiarotti, Cococcioni, Dabo
  et~al.}}]{QE-2009}
\bibinfo{author}{\bibfnamefont{P.}~\bibnamefont{Giannozzi}},
  \bibinfo{author}{\bibfnamefont{S.}~\bibnamefont{Baroni}},
  \bibinfo{author}{\bibfnamefont{N.}~\bibnamefont{Bonini}},
  \bibinfo{author}{\bibfnamefont{M.}~\bibnamefont{Calandra}},
  \bibinfo{author}{\bibfnamefont{R.}~\bibnamefont{Car}},
  \bibinfo{author}{\bibfnamefont{C.}~\bibnamefont{Cavazzoni}},
  \bibinfo{author}{\bibfnamefont{D.}~\bibnamefont{Ceresoli}},
  \bibinfo{author}{\bibfnamefont{G.~L.} \bibnamefont{Chiarotti}},
  \bibinfo{author}{\bibfnamefont{M.}~\bibnamefont{Cococcioni}},
  \bibinfo{author}{\bibfnamefont{I.}~\bibnamefont{Dabo}}, \bibnamefont{et~al.},
  \bibinfo{journal}{J. Phys. Condens. Matter} \textbf{\bibinfo{volume}{21}},
  \bibinfo{pages}{395502} (\bibinfo{year}{2009}),
  \urlprefix\url{http://www.quantum-espresso.org}.

\bibitem[{\citenamefont{Troullier and
  Martins}(1991)}]{TM-PP_Troullier-Martins-PRB-1991}
\bibinfo{author}{\bibfnamefont{N.}~\bibnamefont{Troullier}} \bibnamefont{and}
  \bibinfo{author}{\bibfnamefont{J.~L.} \bibnamefont{Martins}},
  \bibinfo{journal}{Phys. Rev. B} \textbf{\bibinfo{volume}{43}},
  \bibinfo{pages}{1993} (\bibinfo{year}{1991}).

\bibitem[{\citenamefont{Perdew et~al.}(1996)}]{PBE_exch-correl_PRL-1996}
\bibinfo{author}{\bibfnamefont{J.~P.}~\bibnamefont{Perdew}},
  \bibinfo{author}{\bibfnamefont{K.}~\bibnamefont{Burke}}, \bibnamefont{and}
  \bibinfo{author}{\bibfnamefont{M.} \bibnamefont{Ernzerhof}},
  \bibinfo{journal}{Phys. Rev. Lett.} \textbf{\bibinfo{volume}{77}},
  \bibinfo{pages}{3865} (\bibinfo{year}{1996}).

\bibitem[{\citenamefont{Johnson and
  Ashcroft}(2000)}]{metallic_H2_structs_Ashcroft-Nat-2000}
\bibinfo{author}{\bibfnamefont{K.~A.} \bibnamefont{Johnson}} \bibnamefont{and}
  \bibinfo{author}{\bibfnamefont{N.~W.} \bibnamefont{Ashcroft}},
  \bibinfo{journal}{Nature} \textbf{\bibinfo{volume}{403}},
  \bibinfo{pages}{632} (\bibinfo{year}{2000}).

\bibitem[{\citenamefont{Geng et~al.}(2010)\citenamefont{Geng, Li, and
  Wu}}]{triatomic_H_structs_Chinese-arxiv-2010}
\bibinfo{author}{\bibfnamefont{H.~Y.} \bibnamefont{Geng}},
  \bibinfo{author}{\bibfnamefont{J.~F.} \bibnamefont{Li}}, \bibnamefont{and}
  \bibinfo{author}{\bibfnamefont{Q.}~\bibnamefont{Wu}} (\bibinfo{year}{2010}),
  \urlprefix\url{http://arxiv.org/abs/1010.3392}.

\bibitem[{\citenamefont{Pierleoni and
  Ceperley}(2006)}]{CEIMC_Ceperley-LecNotPhys-2006}
\bibinfo{author}{\bibfnamefont{C.}~\bibnamefont{Pierleoni}} \bibnamefont{and}
  \bibinfo{author}{\bibfnamefont{D.~M.} \bibnamefont{Ceperley}},
  \bibinfo{journal}{Lect. Notes Phys.} \textbf{\bibinfo{volume}{703}},
  \bibinfo{pages}{641} (\bibinfo{year}{2006}).

\end{thebibliography}

\end{document}